\documentclass[
  aps,
  prl,
  preprint,
  superscriptaddress,
  floatfix,
]{revtex4-2}

\usepackage{amsmath,amssymb,bm}
\usepackage{graphicx}
\usepackage{xcolor}
\usepackage{hyperref}
\usepackage{float}
\usepackage{soul}
\hypersetup{colorlinks=true, linkcolor=blue, citecolor=blue, urlcolor=blue}

\newcommand{\ii}{\mathrm{i}}
\newcommand{\ee}{\mathrm{e}}

\begin{document}

\title{A microscopic origin for the breakdown of the Stokes–Einstein relation in ion transport}

\author{Zhenyu Wei}
\thanks{These authors contributed equally to this work.}
\author{Mu Chen}
\thanks{These authors contributed equally to this work.}

\author{Jun Ren}
\author{Pinyao He}
\author{Wei Xu}
\author{Wei Liu}
\author{Fei Zheng}
\author{Yin Zhang}
\author{Wei Si}
\author{Jinjie Sha}
\author{Zhonghua Ni}
\author{Yunfei Chen}
\thanks{Corresponding author: yunfeichen@seu.edu.cn}
\affiliation{Jiangsu Key Laboratory for Design and Manufacturing of Precision Medicine Equipment, School of Mechanical Engineering, Southeast University, Nanjing 211189, China}

\begin{abstract}
Ion transport underlies the operation of biological ion channels and governs the performance of electrochemical energy-storage devices. A long-standing anomaly is that smaller alkali metal ions, such as Li$^+$, migrate more slowly in water than larger ions, in apparent violation of the Stokes-Einstein relation. This breakdown is conventionally attributed to dielectric friction, a collective drag force arising from electrostatic interactions between a drifting ion and its surrounding solvent. Here, combining nanopore transport measurements over electric fields spanning several orders of magnitude with molecular dynamics simulations, we show that the time-averaged electrostatic force on a migrating ion is not a drag force but a net driving force. By contrasting charged ions with neutral particles, we reveal that ionic charge introduces additional Lorentzian peaks in the frequency-dependent friction coefficient. These peaks originate predominantly from short-range Lennard-Jones (LJ) interactions within the first hydration layer and represent additional channels for energy dissipation, strongest for Li$^+$ and progressively weaker for Na$^+$ and K$^+$. Our results demonstrate that electrostatic interactions primarily act to tighten the local hydration structure, thereby amplifying short-range LJ interactions rather than directly opposing ion motion. This microscopic mechanism provides a unified physical explanation for the breakdown of the Stokes-Einstein relation in aqueous ion transport.
\end{abstract}

\maketitle

Ion mobility quantifies how rapidly an ion transports through a medium under an applied electric field, revealing the ion selectivity of ion channels\cite{Owsianik2006, Maffeo2012a} and determining the performance of energy-storage devices\cite{Hosaka2020, Quilty2023}. A precise description of ion mobility, however, remains a significant challenge \cite{Wolynes1980,Banerjee2019}. Hydrodynamic theory predicts that an ion in aqueous solution experiences Stokes friction, a force proportional to its radius and the solvent's viscosity. Therefore, mobility, which is inversely proportional to the friction coefficient, should be larger for smaller ions, known as the Stokes-Einstein relation. However, experimental results do not support this prediction, showing that smaller alkali metal ions like Li$^+$ exhibit lower mobilities than their larger counterparts, such as K$^+$. The breakdown of the Stokes-Einstein relationship points to the existence of an additional friction force originating from ion-water interactions.

This additional friction has long been attributed to a collective drag force generated by the electrostatic interactions between a migrating ion and its surrounding water, termed dielectric friction\cite{Born1920}. Continuum models were first developed to describe this effect by treating the solvent as a uniform dielectric medium\cite{Fuoss1959}. The migrating ion polarizes the dielectric medium, and the induced polarization is assumed to decay exponentially in time. The dielectric friction coefficient was then determined from either the energy dissipated in the medium\cite{Boyd1961} or the retarding force acting on the ion from the asymmetric polarization cloud\cite{Zwanzig1963, Zwanzig1970}. However, these models significantly overestimate the dielectric friction coefficient, as the uniform medium assumption cannot capture the distinct structure and dynamics within the ion's hydration layer\cite{Banerjee2019, Biswas1997, Koneshan1998a}. To address these limitations, molecular models were developed that relate the dielectric friction coefficient to the autocorrelation of the electrostatic forces acting on the fixed ion. Wolynes et al.\cite{Colonomos1979, Wolynes1978} assumed that the force autocorrelation decays exponentially with a characteristic time, from which the friction coefficient can be derived. However, the force autocorrelation function of metal alkali metal ions exhibits pronounced fluctuations on timescales far longer than the fitted relaxation time\cite{Berkowitz1987, Nguyen1984, Wilson1985}, making this approximation inadequate for describing dielectric friction in this condition. Without presupposing a specific form for the force autocorrelation, Bagchi et al.\cite{Bagchi1991, Banerjee2017, Biswas1995a, Biswas1997a, Chandra2000a} expressed the electrostatic force as a functional of the solvent’s dipole density distribution. The force autocorrelation function can then be expressed in terms of the dynamic structure factor of the ion and solvent dipoles. The shortcoming of this model lies in its dependence on numerous predefined parameters that must be fitted from experimental or simulation data\cite{Biswas1997a}. Despite these theoretical advances, the underlying mechanism of dielectric friction remains elusive.

\begin{figure}[htbp]
    \begin{center}
        \includegraphics[width=8.6cm]{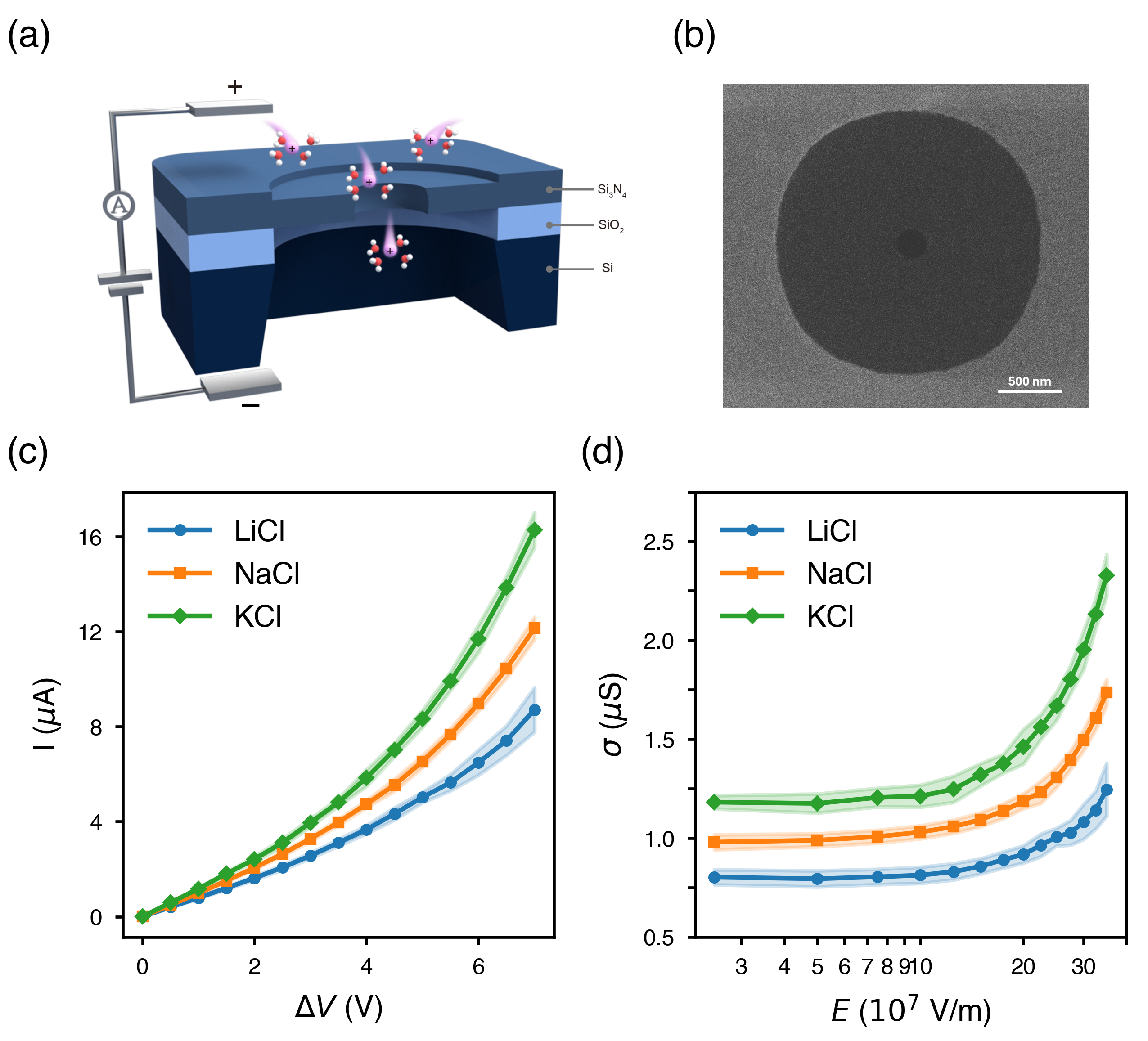}
        \caption{Measuring ion transport through a solid-state nanopore under a wide range electric field intensity. (a), Schematic of the solid-state nanopore system. The membrane has a 20-nm-thick Si$_3$N$_4$ film, with a single nanopore at its center. A bias voltage ($\delta V$) applied across the membrane drives ions through the nanopore, generating an ionic current ($I$). The voltage drop concentrates inside the nanopore, creating an strong electric field ($E$). (b), Scanning electron microscopy (SEM) micrograph of a ~250 nm diameter nanopore. (c), Current-voltage ($I$-$V$) curves for a 250 nm diameter nanopore filled with 0.4 M LiCl, 0.4 M NaCl, and 0.4M KCl solutions, respectively. (d), Nanopore conductance $\sigma$ (derived from data in (c)) as a function of the electric field intensity (E). The conductance remains constant below field 9$\times{10}^7$ V/m and shows a nonlinear increase at higher fields due to Joule heating. Across the entire electrical field range, the conductance follows the order KCl > NaCl > LiCl. In (c) and (d), Data points represent the mean values, and the shaded regions denote standard deviation from 12 independent measurements.}
        \label{fig:experiment-setup}
    \end{center}
\end{figure}

Nanopores with diameters ranging from 50 to 500 nm were fabricated in 20-nm-thick Si$_3$N$_4$ membranes via focused ion beam (FIB) milling to characterize ion transport (Fig. \ref{fig:experiment-setup}(a), (b)). The nanopore membrane is sandwiched between two electrolyte reservoirs. An Ag/AgCl electrode is immersed in each reservoir to apply the bias voltage and recording the resulting ionic current. In comparison with bulk conductivity measurements, which typically employ alternating fields on the order of 10$^2$ V/m, this nanopore setup permits probing ion transport at field strengths spanning 10$^6$-10$^9$ V/m under direct current (DC) conditions. This capability exploits the high resistance of the nanopore, which ensures that the applied bias drops almost entirely across the membrane. Figure \ref{fig:experiment-setup}(c) presents the current-voltage ($I$-$V$) characteristics of a 250-nm-diameter nanopore filled with 0.4 M LiCl, 0.4 M NaCl, and 0.4M KCl solutions, respectively. The corresponding conductance (Fig. \ref{fig:experiment-setup}(d)), derived from the $I$-$V$ characteristics, remains constant at electric field intensities below 10$^8$ V/m. At higher field intensities, the conductance exhibits a nonlinear increment due to Joule heating\cite{Tsutsui2022}. Notably, the conductance ordering KCl > NaCl > LiCl remains invariant across the entire field range. As the anionic contribution (Cl$^-$) is identical, this ordering directly reflects the relative mobilities of the cations, proving that dielectric friction persists and significantly affects ion transport across this wide range of electric fields. Furthermore, this experimental configuration allows for a direct comparison with molecular dynamics (MD) simulations, which require fields on the order of 10$^8$ V/m to resolve an ion's drift velocity from its thermal fluctuations.

\begin{figure}[htbp]
    \begin{center}
        \includegraphics[width=6cm]{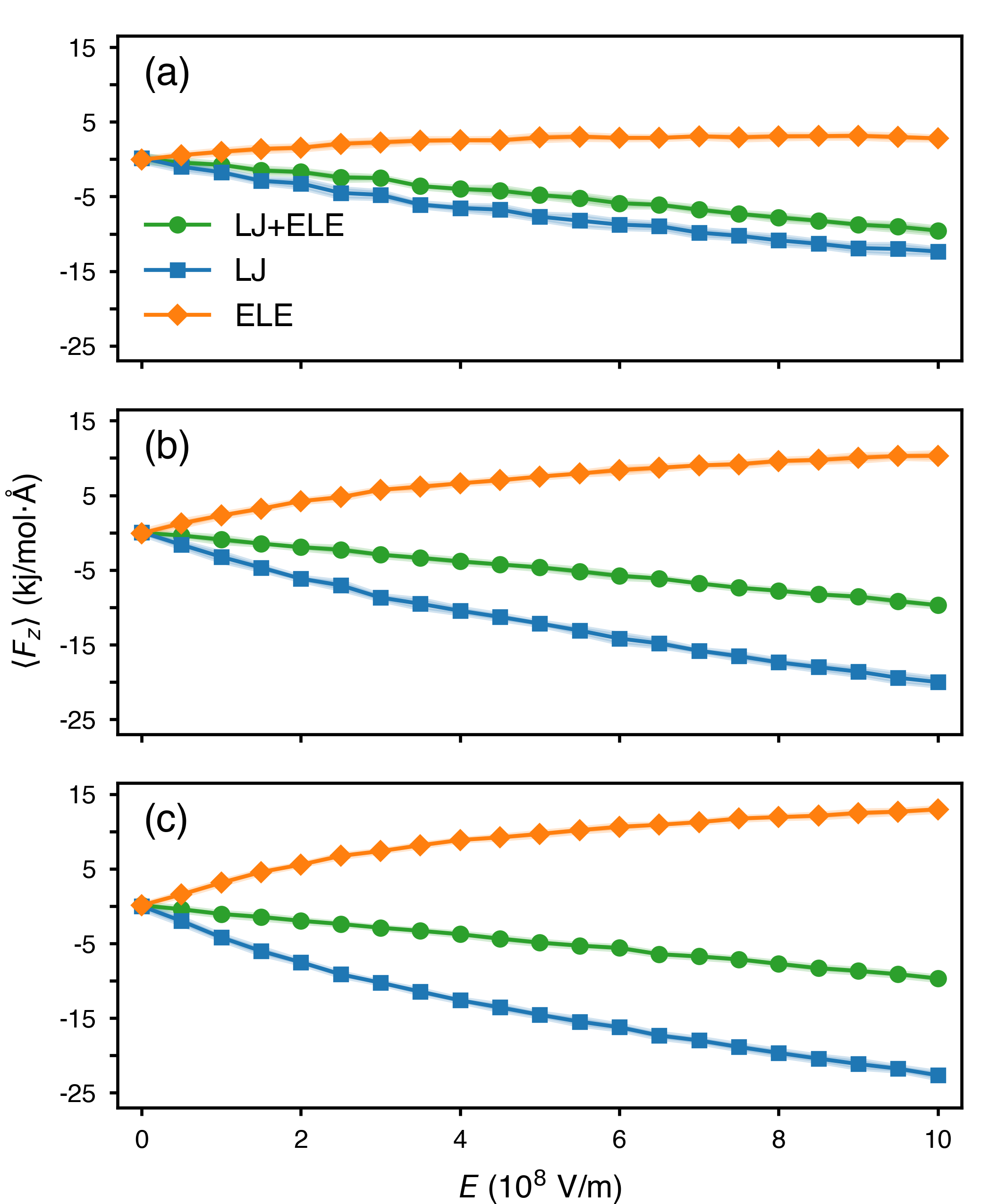}
        \caption{Time averaged force $\left\langle F_z\right\rangle$ acting on ions from ion-water interactions calculated from molecular dynamics (MD) simulation at different applied elect     ric fields. The external electric field was applied along the positive z-direction, and the values plotted represent the force component along this axis. Results are shown  for (a) Li$^+$, (b) Na$^+$, and (c) K$^+$. The total force is decomposed into the contribution from Lennard-Jones (LJ) interaction and the electrostatic (ELE) interaction between ion and water. For all three ions, the ELE component consistently acts as a net driving force, while LJ component acts as a net drag force. Data points represent the mean, and the shaded regions denote standard deviation from 5 independent simulations.}
        \label{fig:averaged-frc}
    \end{center}
\end{figure}

To investigate the microscopic origins of dielectric friction, MD simulations were conducted on a system consisting of a single ion pair solvated in a 40$\times$40$\times$40 \r{A}$^3$ cubic water box (see Sec. S3 of the Supplemental Material for the system illustration and detailed MD setup), thereby minimizing ion-ion interactions. An external electric field on the order of 10$^8$ V/m was applied along the z-axis, and the instantaneous force exerted by the water on the ion was recorded. As shown in Fig. \ref{fig:averaged-frc}, the simulations show that the time-averaged force from ion-water interactions acts as a drag force for all Li$^+$, Na$^+$, and K$^+$ ions. Decomposing this force reveals a fundamental anomaly: the electrostatic (ELE) interaction exerts a net driving force aligned with the drift velocity, whereas the Lennard-Jones (LJ) interaction exerts the drag force. This finding fundamentally contradicts the conventional interpretation of dielectric friction, a collective drag force arising from the electrostatic interaction between the migrating ion and the surrounding solvent. This fundamental anomaly necessitates a reexamination of dielectric friction from first principles.

\begin{figure}[htbp]
    \begin{center}
        \includegraphics[width=16cm]{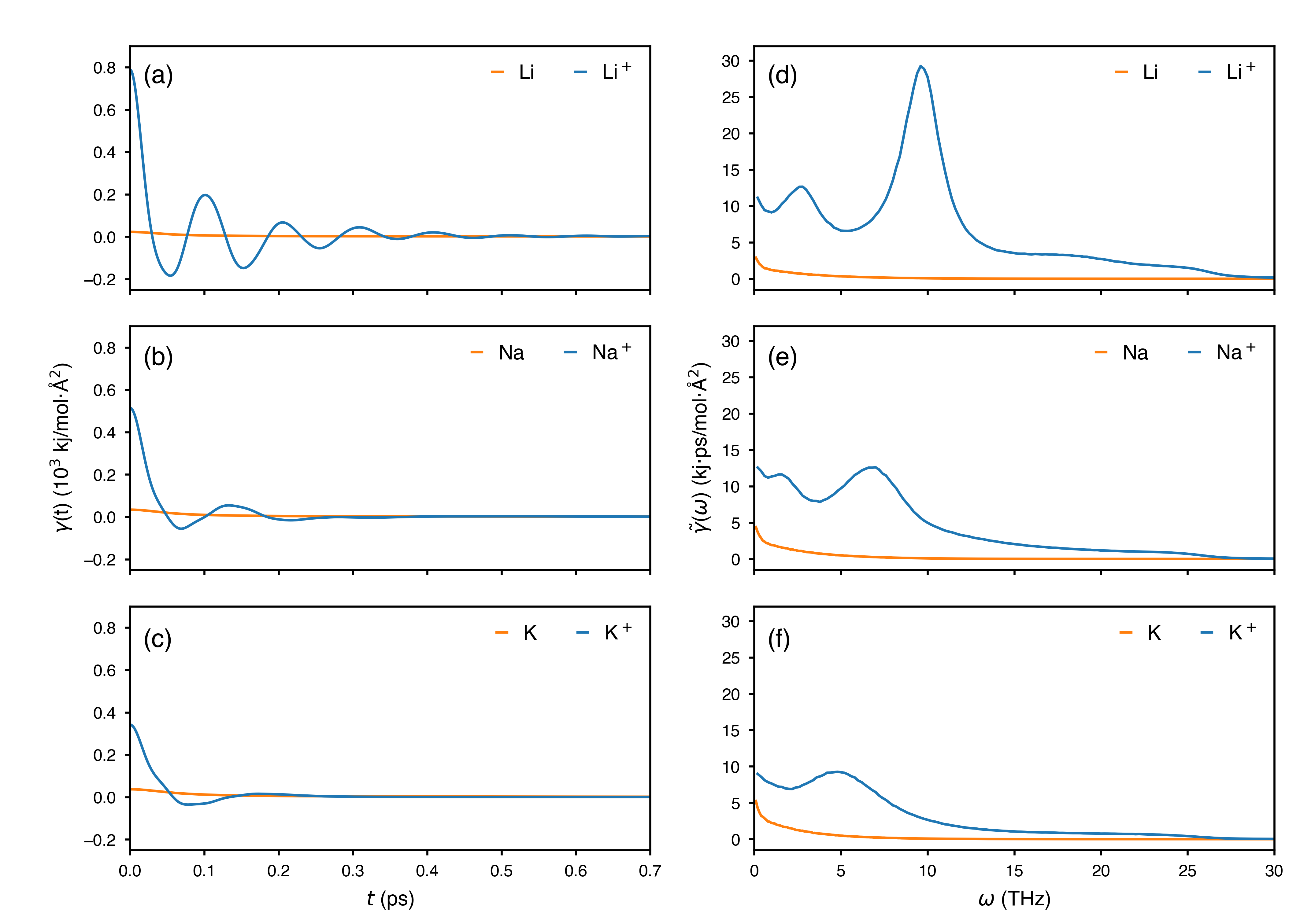}
        \caption{Effect of electrostatic interactions between ion and water on the friction memory kernel and frequency-dependent friction coefficient. (a-c), Friction memory kernel, $\gamma(t)$, for (a) Li$^+$, (b) Na$^+$, and (c) K$^+$. (d-f), Corresponding frequency-dependent friction coefficient, $\tilde{\gamma}(\omega)$, obtained from the Fourier transform of $\gamma(t)$, for (d) Li$^+$, (e) Na$^+$, and (f) K$^+$. In each panel, the result for the standard ion (blue line) is compared with an uncharged counterpart (orange line) where the charge of ion was set to zero to exclude electrostatic interactions. The inclusion of electrostatics introduces significant, decaying fluctuations in the friction memory kernel, most notably for Li$^+$. These fluctuations correspond to the distinct Lorentzian peaks observed in the terahertz regime of the frequency-dependent friction coefficient.}
        \label{fig:charge-comparison}
    \end{center}
\end{figure}

Microscopically, the motion of an ion in aqueous solution is governed by the Hamiltonian equations, with the coordinates and momenta of all particles as variables. Through the Mori-Zwanzig projection, these high-dimensional dynamics can be rigorously reduced to the generalized Langevin equation\cite{Zwanzig2001} (GLE). The detailed derivation can be found in Sec. S6 of the Supplemental Material. In GLE, the ion's velocity is treated as the sole variable, while all remaining degrees of freedom are coarse-grained into an effective heat bath:
\begin{equation}
    m\frac{\mathrm{d}v}{\mathrm{d}t} = -\int_0^\infty \gamma(\tau)v(t-\tau) \mathrm{d}\tau + \delta F(t)
    \label{eq:gle}
\end{equation}
The right-hand side describes the exchange of energy between the ion and the heat bath through fluctuation–dissipation processes. The first term is the friction force, expressed as a convolution of the memory kernel $\gamma(\tau)$ with the ion's velocity history $v(t-\tau)$. This convolution form captures the delayed response of the surrounding medium, with dissipation arising from the relaxation of the heat bath perturbed by the ion's prior motion. The second term, $\delta F\left(t\right)$, denotes the fluctuating force exerted by the heat bath. Energy conservation requires a balance between the dissipation and fluctuation, leading to:
\begin{equation}
    \gamma(\tau) = \frac{\left\langle \delta F(t)\delta F(0)\right\rangle}{k_BT}
    \label{eq:fluctuation-dissipation-theorem}
\end{equation}
The velocity of the ion can be represented in its Fourier expansion form:
\begin{equation}
    v(t) = \int_{-\infty}^{\infty}\frac{\tilde{v}(\omega)}{2\pi} \ee^{\ii\omega t}\mathrm{d}\omega
\end{equation}
Substituting this expression into the friction force term in Eq. \eqref{eq:gle} (detailed derivation can be found in Sec. S7 of the Supplemental Material) yields:
\begin{equation}
    -\int_0^\infty\gamma(\tau)v(t-\tau)\mathrm{d}\tau =
    -\int_{-\infty}^\infty \tilde{\gamma}(\omega) \frac{\tilde{v}(\omega)}{2\pi}\ee^{\ii\omega t}\mathrm{d}\omega
    \label{eq:friction-spectrum}
\end{equation}
where $\tilde{\gamma}(\omega)$ denotes the Fourier transform of the friction memory kernel $\gamma(t)$. Eq. \eqref{eq:friction-spectrum} demonstrates that the convolution of the friction memory kernel with the velocity history in the time domain is equivalent to the product of the instantaneous velocity of a specific frequency component and $\tilde{\gamma}(\omega)$. $\tilde{\gamma}(\omega)$ represents the frequency-dependent friction coefficient, or friction spectrum, quantifying the effective drag experienced by an ion undergoing harmonic oscillation at frequency $\omega$.

\begin{figure*}[htbp]
  \centering
  \includegraphics[width=16cm]{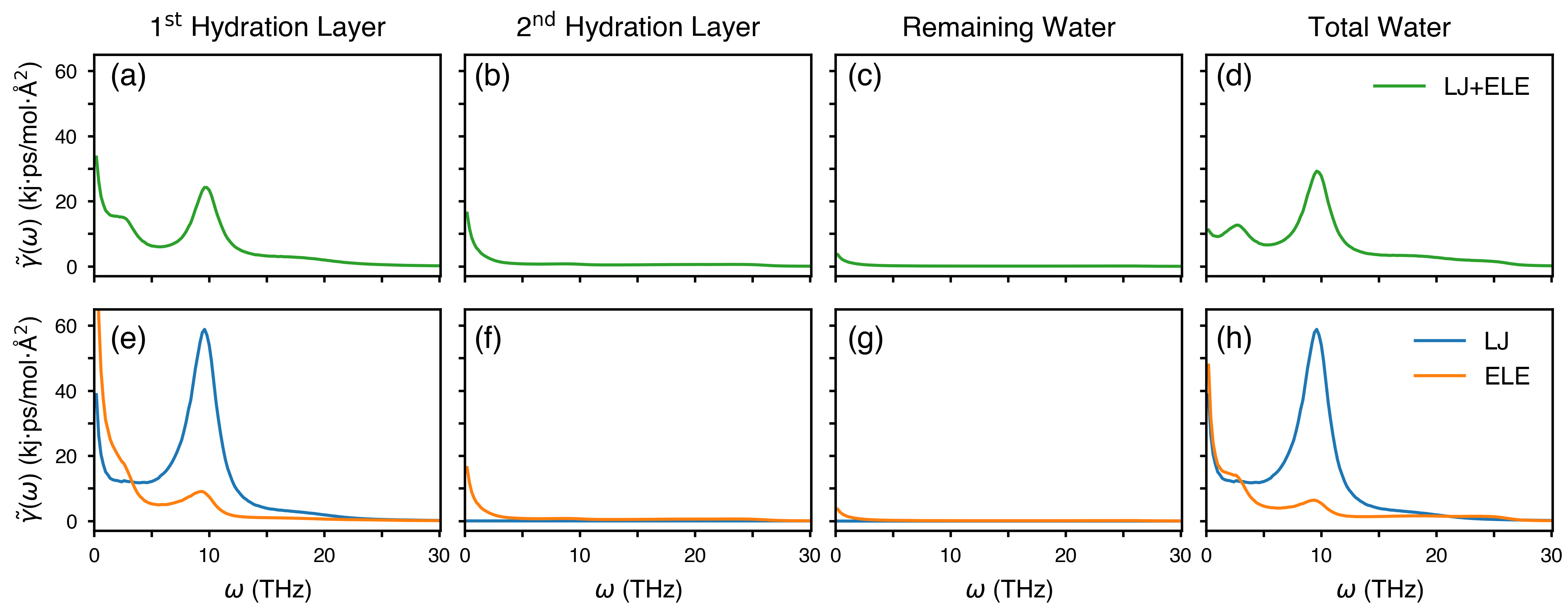}
  \caption{
    Decomposition of the Li$^+$  frequency-dependent friction coefficient  by spatial region and interaction type. (a-d), Spatially decomposed frequency-dependent friction coefficient, $\tilde{\gamma}(\omega)$ , showing the contribution from interaction between ion and (a) 1$^\text{st}$ hydration layer, (b) 2$^\text{nd}$ hydration layer, (c) remaining bulk solvent, and (d) the total water environment. (e-h), Further decomposition of each spatial component into its Lennard-Jones (LJ, blue) and electrostatic (ELE, orange) contributions: (e) 1$^\text{st}$ hydration layer, (f) 2$^\text{nt}$ hydration layer, (g) remaining bulk solvent, and (h) the total water environment.
  }
  \label{fig:friction-spectrum-decomposition}
\end{figure*}

Based on Eq. \eqref{eq:fluctuation-dissipation-theorem}, we performed additional MD simulations on the same system, but with the ion held fixed, to extract the friction memory kernel and the corresponding frequency-dependent friction coefficient. In this fixed-ion configuration, the ion velocity vanishes, eliminating the friction term in the generalized Langevin equation. As a result, the instantaneous force acting on the stationary ion is entirely given by the fluctuating force, $\delta F(t)$. Assuming that this fluctuating force retains the same statistical properties as that acting on a mobile ion, the friction memory kernel for the mobile ion can be obtained directly from the autocorrelation function of the instantaneous force experienced by the fixed ion via the fluctuation-dissipation theorem.

To isolate the contribution of electrostatic interactions between an ion and its surrounding molecules to dielectric friction, we set the ion's charge to zero and computed the resulting friction memory kernel, which we then compared with that of the fully charged ion (Fig. \ref{fig:charge-comparison}(a)-(c)). In the absence of charge, the friction memory kernel decays exponentially. By contrast, the kernel exhibits pronounced fluctuations for charged ions, most prominently for Li$^+$. The corresponding friction spectra (Fig. \ref{fig:charge-comparison}(d)-(f)) reveal that, beyond the Debye peak centered at $\omega=0$ present in both charged and neutral cases, ionic charge induces distinct Lorentzian peaks in the terahertz regime, representing additional energy dissipation channels. Notably, the Lorentzian peak near 10 THz in the spectrum of Li$^+$ (Fig. \ref{fig:charge-comparison}(d)) has the largest magnitude, reflecting the strongest energy dissipation, whereas the corresponding peak near 5 THz in the K$^+$ spectrum (Fig. \ref{fig:charge-comparison}(f)) is substantially weaker. This trend demonstrates that the strength of these charge-induced dissipation channels increases with decreasing ionic size. Importantly, this behavior is incompatible with conventional dielectric-friction models that attribute dissipation primarily to collective electrostatic drag. It is also distinct from earlier studies of dielectric friction associated with vibrational relaxation of covalent bonds in molecular systems such as CH$_3$Cl or HgI$_2$\cite{Bruehl1993, Gnanakaran1996, Ladanyi1999, Whitnell1990, Whitnell1992}.

To elucidate the microscopic origin of the additional dissipation channels, we decomposed the force arising from the total ion-water interaction into contributions from the first hydration layer, the second hydration layer, and the remaining bulk water. The corresponding friction spectra for each region are shown in Fig. \ref{fig:friction-spectrum-decomposition}(a)-(d). The spectrum associated with the first hydration layer (Fig. \ref{fig:friction-spectrum-decomposition}(a)) exhibits pronounced Lorentzian peaks in the terahertz regime, in addition to the Debye peak at zero frequency. By contrast, the spectra arising from the second hydration layer (Fig. \ref{fig:friction-spectrum-decomposition}(b)) and the remaining bulk water (Fig. \ref{fig:friction-spectrum-decomposition}(c)) display only a Debye peak, with the peak of second layer exhibiting a larger amplitude than that of remaining bulk water. These results demonstrate that the additional dissipation channel, manifested as Lorentzian features in the total friction spectrum (Fig. \ref{fig:friction-spectrum-decomposition}(d)), originates exclusively from ion-water interactions within the first hydration layer.

We further decomposed the friction spectra in each spatial region into contributions from LJ and electrostatic interactions (Fig. \ref{fig:friction-spectrum-decomposition}(e)-(h)). This decomposition shows that the friction spectra associated with the second hydration layer (Fig. \ref{fig:friction-spectrum-decomposition}(f)) and the remaining bulk water (Fig. \ref{fig:friction-spectrum-decomposition}(g)) are dominated by electrostatic interactions. In sharp contrast, the first hydration layer exhibits different behavior: the prominent Lorentzian peak near 10 THz in its spectrum (Fig. \ref{fig:friction-spectrum-decomposition}(e)) is overwhelmingly governed by LJ interactions, with electrostatic contributions being substantially smaller. These results demonstrate that the additional energy dissipation channel originates primarily from short-range LJ interactions between the ion and the tightly bound water molecules in the first hydration layer (see Sec. S5 of the Supplemental Material), rather than from direct electrostatic drag.

In summary, our experiments show that dielectric friction persists and significantly affects ion transport across electric-field strengths spanning 10$^6$ to 10$^8$ V/m, leading to the anomalously low mobility of small alkali metal ions such as Li$^+$ relative to Na$^+$ and K$^+$. Molecular dynamics simulations further reveal that the time-averaged electrostatic force between a migrating ion and its hydration environment acts as a driving force rather than a drag force, challenging the long-standing view that dielectric friction is a collective drag force arising from the electrostatic interaction. Instead, the primary role of electrostatic interactions is to tighten the ion's first hydration layer. This enhanced structuring strengthens the LJ interaction between the ion and surrounding water molecules, introducing additional energy-dissipation channels that appear as Lorentzian peaks in the terahertz regime of the frequency-dependent friction coefficient. These dissipation channels constitute the microscopic origin of dielectric friction and offer a mechanistic explanation for the breakdown of the Stokes-Einstein relation for small alkali metal ions in aqueous solutions.

\begin{acknowledgments}
The authors thank the Natural Science Foundation of China (Grants No. 52127811, T2321002, 52441506) and Natural Science Foundation of Jiangsu Province, Major Project (BK20222005). The authors thank the Center for Fundamental and Interdisciplinary Sciences of Southeast University for the support in friction phonon detection and measurement. This research work is supported by the Big Data Computing Center of Southeast University.
\end{acknowledgments}


\bibliographystyle{apsrev4-2}
\bibliography{ref}

@article{Bagchi1991,
  title = {Microscopic Expression for Dielectric Friction on a Moving Ion},
  author = {Bagchi, Biman},
  year = 1991,
  month = jul,
  journal = {The Journal of Chemical Physics},
  volume = {95},
  number = {1},
  pages = {467--478},
  issn = {0021-9606, 1089-7690},
  doi = {10.1063/1.461447},
  urldate = {2024-01-06},
  langid = {english},
  lccn = {4.304}
}

@article{Banerjee2017,
  title = {A Mode-Coupling Theory Analysis of the Observed Diffusion Anomaly in Aqueous Polyatomic Ions},
  author = {Banerjee, Puja and Bagchi, Biman},
  year = 2017,
  month = sep,
  journal = {The Journal of Chemical Physics},
  volume = {147},
  number = {12},
  pages = {124502},
  issn = {0021-9606, 1089-7690},
  doi = {10.1063/1.4994631},
  urldate = {2024-01-05},
  langid = {english},
  lccn = {4.304}
}

@article{Banerjee2019,
  title = {Ions' Motion in Water},
  author = {Banerjee, Puja and Bagchi, Biman},
  year = 2019,
  month = may,
  journal = {The Journal of Chemical Physics},
  volume = {150},
  number = {19},
  pages = {190901},
  issn = {0021-9606, 1089-7690},
  doi = {10.1063/1.5090765},
  urldate = {2023-07-18},
  langid = {english},
  lccn = {4.304}
}

@article{Berkowitz1987,
  title = {The Limiting Ionic Conductivity of Na+ and Cl- Ions in Aqueous Solutions: Molecular Dynamics Simulation},
  shorttitle = {The Limiting Ionic Conductivity of Na+ and Cl- Ions in Aqueous Solutions},
  author = {Berkowitz, Max and Wan, W.},
  year = 1987,
  month = jan,
  journal = {The Journal of Chemical Physics},
  volume = {86},
  number = {1},
  pages = {376--382},
  issn = {0021-9606, 1089-7690},
  doi = {10.1063/1.452574},
  urldate = {2024-02-26},
  langid = {english},
  lccn = {4.304}
}

@article{Biswas1995a,
  title = {Anomalous Ion Diffusion in Dense Dipolar Liquids},
  author = {Biswas, Ranjit and Roy, Srabani and Bagchi, Biman},
  year = 1995,
  month = aug,
  journal = {Physical Review Letters},
  volume = {75},
  number = {6},
  pages = {1098--1101},
  issn = {0031-9007, 1079-7114},
  doi = {10.1103/PhysRevLett.75.1098},
  urldate = {2025-10-02},
  copyright = {http://link.aps.org/licenses/aps-default-license},
  langid = {english}
}

@article{Biswas1997,
  title = {Ionic Mobility in Alcohols: From Dielectric Friction to the Solvent--Berg Model},
  shorttitle = {Ionic Mobility in Alcohols},
  author = {Biswas, Ranjit and Bagchi, Biman},
  year = 1997,
  month = apr,
  journal = {The Journal of Chemical Physics},
  volume = {106},
  number = {13},
  pages = {5587--5598},
  issn = {0021-9606, 1089-7690},
  doi = {10.1063/1.473581},
  urldate = {2025-10-02},
  langid = {english}
}

@article{Biswas1997a,
  title = {Limiting Ionic Conductance of Symmetrical, Rigid Ions in Aqueous Solutions: Temperature Dependence and Solvent Isotope Effects},
  shorttitle = {Limiting Ionic Conductance of Symmetrical, Rigid Ions in Aqueous Solutions},
  author = {Biswas, Ranjit and Bagchi, Biman},
  year = 1997,
  month = jun,
  journal = {Journal of the American Chemical Society},
  volume = {119},
  number = {25},
  pages = {5946--5953},
  issn = {0002-7863, 1520-5126},
  doi = {10.1021/ja970118o},
  urldate = {2025-10-02},
  langid = {english}
}

@article{Born1920,
  title = {\"Uber Die Beweglichkeit Der Elektrolytischen Ionen},
  author = {Born, M.},
  year = 1920,
  month = jun,
  journal = {Zeitschrift f\"ur Physik},
  volume = {1},
  number = {3},
  pages = {221--249},
  issn = {1434-6001},
  doi = {10.1007/BF01329168},
  urldate = {2024-04-29},
  copyright = {http://www.springer.com/tdm},
  langid = {english}
}

@article{Boyd1961,
  title = {Extension of Stokes' Law for Ionic Motion to Include the Effect of Dielectric Relaxation},
  author = {Boyd, Richard H.},
  year = 1961,
  month = oct,
  journal = {The Journal of Chemical Physics},
  volume = {35},
  number = {4},
  pages = {1281--1283},
  issn = {0021-9606, 1089-7690},
  doi = {10.1063/1.1732039},
  urldate = {2024-02-26},
  langid = {english},
  lccn = {4.304}
}

@article{Bruehl1993,
  title = {Vibrational Relaxation Times for a Model Hydrogen-Bonded Complex in a Polar Solvent},
  author = {Bruehl, Margaret and Hynes, James T.},
  year = 1993,
  month = sep,
  journal = {Chemical Physics},
  volume = {175},
  number = {1},
  pages = {205--221},
  issn = {03010104},
  doi = {10.1016/0301-0104(93)80238-5},
  urldate = {2025-11-29},
  copyright = {https://www.elsevier.com/tdm/userlicense/1.0/},
  langid = {english}
}

@article{Chandra2000a,
  title = {Frequency Dependence of Ionic Conductivity of Electrolyte Solutions},
  author = {Chandra, Amalendu and Bagchi, Biman},
  year = 2000,
  month = jan,
  journal = {The Journal of Chemical Physics},
  volume = {112},
  number = {4},
  pages = {1876--1886},
  issn = {0021-9606, 1089-7690},
  doi = {10.1063/1.480751},
  urldate = {2025-08-27},
  langid = {english}
}

@article{Colonomos1979,
  title = {Molecular Theory of Solvated Ion Dynamics. II. Fluid Structure and Ionic Mobilities},
  author = {Colonomos, Peter and Wolynes, Peter G.},
  year = 1979,
  month = sep,
  journal = {The Journal of Chemical Physics},
  volume = {71},
  number = {6},
  pages = {2644--2651},
  issn = {0021-9606, 1089-7690},
  doi = {10.1063/1.438621},
  urldate = {2024-01-08},
  langid = {english},
  lccn = {4.304}
}

@article{Fuoss1959,
  title = {Dependence of the Walden Product on Dielectric Constant},
  author = {Fuoss, Raymond M.},
  year = 1959,
  month = jun,
  journal = {Proceedings of the National Academy of Sciences},
  volume = {45},
  number = {6},
  pages = {807--813},
  issn = {0027-8424},
  doi = {10.1073/pnas.45.6.807},
  urldate = {2024-03-28},
  langid = {english},
  lccn = {1}
}

@article{Gnanakaran1996,
  title = {Vibrational Relaxation of HgI in Ethanol: Equilibrium Molecular Dynamics Simulations},
  shorttitle = {Vibrational Relaxation of HgI in Ethanol},
  author = {Gnanakaran, S. and Hochstrasser, R. M.},
  year = 1996,
  month = sep,
  journal = {The Journal of Chemical Physics},
  volume = {105},
  number = {9},
  pages = {3486--3496},
  issn = {0021-9606, 1089-7690},
  doi = {10.1063/1.472218},
  urldate = {2025-11-29},
  langid = {english}
}

@article{Hosaka2020,
  title = {Research Development on K-Ion Batteries},
  author = {Hosaka, Tomooki and Kubota, Kei and Hameed, A. Shahul and Komaba, Shinichi},
  year = 2020,
  month = jul,
  journal = {Chemical Reviews},
  volume = {120},
  number = {14},
  pages = {6358--6466},
  issn = {0009-2665, 1520-6890},
  doi = {10.1021/acs.chemrev.9b00463},
  urldate = {2025-11-27},
  copyright = {https://doi.org/10.15223/policy-029},
  langid = {english}
}

@article{Koneshan1998a,
  title = {Solvent Structure, Dynamics, and Ion Mobility in Aqueous Solutions at 25 {$^\circ$}C},
  author = {Koneshan, S. and Rasaiah, Jayendran C. and {Lynden-Bell}, R. M. and Lee, S. H.},
  year = 1998,
  month = may,
  journal = {The Journal of Physical Chemistry B},
  volume = {102},
  number = {21},
  pages = {4193--4204},
  issn = {1520-6106, 1520-5207},
  doi = {10.1021/jp980642x},
  urldate = {2025-10-02},
  langid = {english}
}

@article{Ladanyi1999,
  title = {On the Role of Dielectric Friction in Vibrational Energy Relaxation},
  author = {Ladanyi, Branka M. and Stratt, Richard M.},
  year = 1999,
  month = aug,
  journal = {The Journal of Chemical Physics},
  volume = {111},
  number = {5},
  pages = {2008--2018},
  issn = {0021-9606, 1089-7690},
  doi = {10.1063/1.479469},
  urldate = {2025-08-30},
  langid = {english}
}

@article{Maffeo2012a,
  title = {Modeling and Simulation of Ion Channels},
  author = {Maffeo, Christopher and Bhattacharya, Swati and Yoo, Jejoong and Wells, David and Aksimentiev, Aleksei},
  year = 2012,
  month = dec,
  journal = {Chemical Reviews},
  volume = {112},
  number = {12},
  pages = {6250--6284},
  issn = {0009-2665, 1520-6890},
  doi = {10.1021/cr3002609},
  urldate = {2025-11-27},
  langid = {english}
}

@article{Nguyen1984,
  title = {Studies of Solvated Ion Motion: Molecular Dynamics Results for Dilute Aqueous Solutions of Alkali and Halide Ions},
  shorttitle = {Studies of Solvated Ion Motion},
  author = {Nguyen, H. L. and Adelman, Steven A.},
  year = 1984,
  month = nov,
  journal = {The Journal of Chemical Physics},
  volume = {81},
  number = {10},
  pages = {4564--4573},
  issn = {0021-9606, 1089-7690},
  doi = {10.1063/1.447430},
  urldate = {2024-02-25},
  langid = {english},
  lccn = {4.304}
}

@article{Owsianik2006,
  title = {Permeation and Selectivity of Trp Channels},
  author = {Owsianik, Grzegorz and Talavera, Karel and Voets, Thomas and Nilius, Bernd},
  year = 2006,
  month = jan,
  journal = {Annual Review of Physiology},
  volume = {68},
  number = {1},
  pages = {685--717},
  issn = {0066-4278, 1545-1585},
  doi = {10.1146/annurev.physiol.68.040204.101406},
  urldate = {2025-11-27},
  langid = {english}
}

@article{Quilty2023,
  title = {Electron and Ion Transport in Lithium and Lithium-Ion Battery Negative and Positive Composite Electrodes},
  author = {Quilty, Calvin D. and Wu, Daren and Li, Wenzao and Bock, David C. and Wang, Lei and Housel, Lisa M. and Abraham, Alyson and Takeuchi, Kenneth J. and Marschilok, Amy C. and Takeuchi, Esther S.},
  year = 2023,
  month = feb,
  journal = {Chemical Reviews},
  volume = {123},
  number = {4},
  pages = {1327--1363},
  issn = {0009-2665, 1520-6890},
  doi = {10.1021/acs.chemrev.2c00214},
  urldate = {2025-11-27},
  copyright = {https://doi.org/10.15223/policy-029},
  langid = {english}
}

@article{Tsutsui2022,
  title = {Ionic Heat Dissipation in Solid-State Pores},
  author = {Tsutsui, Makusu and Arima, Akihide and Yokota, Kazumichi and Baba, Yoshinobu and Kawai, Tomoji},
  year = 2022,
  month = feb,
  journal = {Science Advances},
  volume = {8},
  number = {6},
  pages = {eabl7002},
  issn = {2375-2548},
  doi = {10.1126/sciadv.abl7002},
  urldate = {2023-01-06},
  langid = {english},
  lccn = {1}
}

@article{Whitnell1990,
  title = {Fast Vibrational Relaxation for a Dipolar Molecule in a Polar Solvent},
  author = {Whitnell, Robert M. and Wilson, Kent R. and Hynes, James T.},
  year = 1990,
  month = nov,
  journal = {The Journal of Physical Chemistry},
  volume = {94},
  number = {24},
  pages = {8625--8628},
  issn = {0022-3654, 1541-5740},
  doi = {10.1021/j100387a002},
  urldate = {2025-11-29},
  langid = {english}
}

@article{Whitnell1992,
  title = {Vibrational Relaxation of a Dipolar Molecule in Water},
  author = {Whitnell, Robert M. and Wilson, Kent R. and Hynes, James T.},
  year = 1992,
  month = apr,
  journal = {The Journal of Chemical Physics},
  volume = {96},
  number = {7},
  pages = {5354--5369},
  issn = {0021-9606, 1089-7690},
  doi = {10.1063/1.462720},
  urldate = {2025-11-29},
  langid = {english}
}

@article{Wilson1985,
  title = {Molecular Dynamics Test of the Brownian Description of Na+ Motion in Water},
  author = {Wilson, Michael A. and Pohorille, Andrew and Pratt, Lawrence R.},
  year = 1985,
  month = dec,
  journal = {The Journal of Chemical Physics},
  volume = {83},
  number = {11},
  pages = {5832--5836},
  issn = {0021-9606, 1089-7690},
  doi = {10.1063/1.449663},
  urldate = {2024-02-26},
  langid = {english},
  lccn = {4.304}
}

@article{Wolynes1978,
  title = {Molecular Theory of Solvated Ion Dynamics},
  author = {Wolynes, Peter G.},
  year = 1978,
  month = jan,
  journal = {The Journal of Chemical Physics},
  volume = {68},
  number = {2},
  pages = {473--483},
  issn = {0021-9606, 1089-7690},
  doi = {10.1063/1.435777},
  urldate = {2024-02-26},
  langid = {english},
  lccn = {4.304}
}

@article{Wolynes1980,
  title = {Dynamics of Electrolyte Solutions},
  author = {Wolynes, P G},
  year = 1980,
  month = oct,
  journal = {Annual Review of Physical Chemistry},
  volume = {31},
  number = {1},
  pages = {345--376},
  issn = {0066-426X, 1545-1593},
  doi = {10.1146/annurev.pc.31.100180.002021},
  urldate = {2024-02-25},
  langid = {english}
}

@article{Zwanzig1963,
  title = {Dielectric Friction on a Moving Ion},
  author = {Zwanzig, Robert},
  year = 1963,
  month = apr,
  journal = {The Journal of Chemical Physics},
  volume = {38},
  number = {7},
  pages = {1603--1605},
  issn = {0021-9606, 1089-7690},
  doi = {10.1063/1.1776929},
  urldate = {2023-07-18},
  langid = {english},
  lccn = {4.304}
}

@article{Zwanzig1970,
  title = {Dielectric Friction on a Moving Ion. II. Revised Theory},
  author = {Zwanzig, Robert},
  year = 1970,
  month = apr,
  journal = {The Journal of Chemical Physics},
  volume = {52},
  number = {7},
  pages = {3625--3628},
  issn = {0021-9606, 1089-7690},
  doi = {10.1063/1.1673535},
  urldate = {2023-10-06},
  langid = {english},
  lccn = {4.304}
}

@book{Zwanzig2001,
  title = {Nonequilibrium Statistical Mechanics},
  author = {Zwanzig, R. W.},
  year = 2001,
  publisher = {Oxford university press},
  address = {Oxford New York},
  isbn = {978-0-19-514018-7},
  langid = {english},
  lccn = {530.13}
}

\end{document}